\documentclass[11pt]{article}

\usepackage[a4paper, margin=2.5cm]{geometry}

\usepackage{amsmath}
\usepackage{amsfonts}
\usepackage{dsfont}
\usepackage[T1]{fontenc}

\usepackage{graphicx}
\graphicspath{{./fig/}}

\usepackage[style=authoryear,natbib=true,firstinits=true,sorting=nyt,maxbibnames=1,maxcitenames=1,url=false,backend=bibtex,dashed=false]{biblatex}
\usepackage{fancyhdr}
\usepackage{lineno}

\usepackage{hyperref}
\hypersetup
{
  bookmarksopen=true,
  pdftitle=Time-of-flight proton CT,
  pdfauthor=NilsKrah, 
  pdfmenubar=true, 
  pdfhighlight=/O, 
  colorlinks=true, 
  pdfpagemode=None, 
  pdfpagelayout=SinglePage, 
  pdffitwindow=true, 
  linkcolor=blue, 
  citecolor=blue, 
  urlcolor=blue 
}

\usepackage{xcolor}

\colorlet{jmcolor}{green!70!black}

\usepackage[color=orange,linecolor=orange,textsize=tiny]{todonotes}

\usepackage{indentfirst}
\newcommand*\rmd{\mathop{}\!\mathrm{d}}

\begin{document}
\thispagestyle{empty}
\suppressfloats

\vspace{1em}

\noindent{\huge Relative stopping power precision \\in time-of-flight proton CT}

\vspace{3em}

\noindent Nils Krah$^{1,2}$, Denis Dauvergne$^{3}$, Jean Michel L\'etang$^{1}$, Simon Rit$^{1}$, \'Etienne Testa$^{2}$
\vspace{1em}

\noindent {\scriptsize
$^1$University of Lyon, CREATIS, CNRS UMR5220, Inserm U1044, INSA-Lyon, Universit\'e Lyon 1, Centre L\'eon B\'erard, France \\
$^2$University of Lyon, Universit\'e Claude Bernard Lyon 1, CNRS/IN2P3, IP2I Lyon, UMR 5822, Villeurbanne, France \\
$^3$Universit\'e Grenoble Alpes, CNRS/IN2P3, Grenoble INP, LPSC-UMR 5821, Grenoble, France \\
\emph{nils.krah@creatis.insa-lyon.fr}
}

\begin{abstract}
\mbox{}\par\vspace{-\baselineskip}
\paragraph{Objective}
Proton computed tomography (CT) is similar to x-ray CT but relies on protons rather than photons to form an image. In its most common operation mode, the measured quantity is the amount of energy that a proton has lost while traversing the imaged object from which a relative stopping power map can be obtained via tomographic reconstruction. To this end, a calorimeter which measures the energy deposited by protons downstream of the scanned object
has been studied or implemented as energy detector in several proton CT prototypes. An alternative method is to measure the proton's residual velocity and thus its kinetic energy via the time of flight (TOF) between at least two sensor planes. 
In this work, we study the precision, i.e. image noise, which can be expected from TOF proton CT systems. 

\paragraph{Approach}
We rely on physics models on the one hand and statistical models of the relevant uncertainties on the other to derive closed form expressions for the noise in projection images. The TOF measurement error scales with the distance between the TOF sensor planes and is reported as velocity error in ps/m. We use variance reconstruction to obtain noise maps of a water cylinder phantom given the scanner characteristics and additionally reconstruct noise maps for a calorimeter-based proton CT system as reference.

\paragraph{Main results}
We find that TOF proton CT with 30\,ps/m velocity error
reaches similar image noise as a calorimeter-based proton CT system with 1\% energy error (1 sigma error).  
A TOF proton CT system with a 50\,ps/m velocity error produces slightly less noise than a 2\% calorimeter system. Noise in a reconstructed TOF proton CT image is spatially inhomogeneous with a marked increase towards the object periphery.

\paragraph{Significance} 
This systematic study of image noise in TOF proton CT can serve as a guide for future developments of this alternative solution for estimating the residual energy of protons after the scanned object.

\end{abstract}

\section{Introduction}
Proton computed tomography (CT) is a transmission imaging modality similar to x-ray CT which uses protons instead of x-ray for image acquisition \parencite{Johnson2018}. While x-ray CT relies on attenuation, proton CT is typically operated in energy-loss mode where the amount of energy lost by a proton while traversing the imaged object depends on the water equivalent amount of material. The reconstructed quantity in this case is the so-called relative stopping power (RSP), i.e. the proton  stopping power relative to that of water. A proton CT scanner typically includes a suitable detector to measure the protons' residual energy, i.e. after traversing the scanned object, from which, together with the beam energy, the energy loss can be obtained. Proton CT scanner prototypes proposed and developed so far rely either on a calorimeter, e.g. a scintillator, to measure the energy deposited by a proton therein \parencite{Civinini2013,Johnson2016,DeJongh2020} or a range telescope \parencite{Taylor2016,Alme2020,Pemler1999}. An alternative measurement principle determines the proton's time of flight (TOF) between two or more sensor planes from which the velocity and thus the kinetic energy can be calculated. Although \textcite{Worstell2019} reported on developments of a TOF proton radiography system and presented some first experimental results, TOF proton CT has overall not been fully explored yet. 
\textcite{Ulrich-Pur2021} recently investigated the feasibility of a TOF proton CT system using low gain avalanche detectors \parencite{Pellegrini2014} based on Monte Carlo simulations. They studied RSP accuracy and RSP precision with a focus on the detector hardware and possible calibration procedures. \textcite{Volz2020thesis} discusses TOF as an alternative measurement principle in helium CT and analyses aspects related to RSP resolution. RSP precision has previously been investigated for energy-loss proton CT based on calorimeter detectors \parencite{Schulte2005,Collins-Fekete2021,Radler2018,Dickmann2019} as well as for other proton CT modalities which do not require residual energy measurements \parencite{Quinones2016,Krah2020a,Bopp2015}. 

This work provides a concise and systematic study of RSP precision in TOF proton CT, which manifests itself as image noise, based on physics models and statistics. 
As sources of uncertainty, we consider the TOF measurement error, energy straggling inside the imaged object, as well as the proton beam's energy spread. 
We analyse noise on a projection level and also 
reconstruct noise images \parencite{Wunderlich2008,Radler2018} of a water cylinder to assess the RSP precision depending on the location in the reconstructed image. 
In this context, we consider only direct reconstruction algorithms \parencite{Khellaf2020} and distance driven binning backprojection in particular \parencite{Rit2013}.

The contribution of this work is to provide the necessary tools to quickly model RSP in a TOF proton CT system and to derive figures of merit of the achievable RSP precision given the system's characteristics and the properties of the imaged object, similar to the pioneering work of \textcite{Schulte2005} for energy-loss proton CT. 

\section{Materials and Methods}

\subsection{Statistical aspects of TOF proton CT}\label{sec:tofstraggling}
With noise in the proton CT images, we intend the variation of RSP value which one would observe in each pixel in repeated acquisitions of the same object, as described e.g. in \parencite{Wunderlich2008}. Image noise is therefore synonymous to the RSP precision or RSP resolution. 
The RSP value is obtained via tomographic image reconstruction which takes a series of projections as input. The projections contain water equivalent path length (WEPL) values which are in turn converted from the proton's energy-loss, e.g. via a look-up table. Therefore, statistical errors in the energy-loss are propagated into the reconstructed RSP value. 

\begin{figure}
\centering
\includegraphics[width=0.7\textwidth]{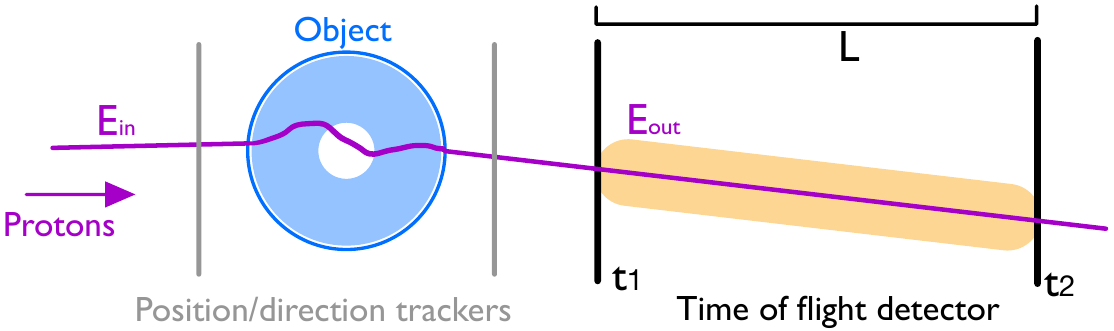}
\caption{Schematic illustration of a TOF proton CT scanner. The residual energy is deduced from the time $t_2-t_1$ which the proton needs to travel from one sensor plane to the other. The trackers which measure the proton's position and direction are shown for completeness, although not directly relevant for this work.}
\label{fig:illustrationscanner}
\end{figure}

We consider a TOF proton CT set-up made of a TOF device consisting of two sensor planes downstream of the scanned object. A proton CT scanner typically also includes tracking devices upstream and downstream of the scanned object, but they are not directly relevant to this work. A sketch is shown in figure~\ref{fig:illustrationscanner}. We assume that the TOF device records the time needed to travel from one sensor plane to the other, from which the proton's kinetic energy downstream of the imaged object can be determined. The initial proton energy is taken as the beam energy provided by the accelerator. We do not attempt to simulate any specific TOF proton CT prototype.

We consider three contributions to image noise, namely the error of the TOF measurement, energy straggling experienced by the protons, and the energy spread of the proton beam. We denote with $E_\mathrm{in}$ and $E_\mathrm{out}$ the proton's energy upstream and downstream of the imaged object, respectively.  
We disregard nuclear interactions and assume that protons recorded by the proton CT system have only undergone electromagnetic interactions. 
In practice, most protons which have undergone nuclear interactions are filtered out from the data prior to image reconstruction \parencite{Johnson2018,Schulte2008}. 
We treat energy straggling and the time measurement error of the TOF device within the Gaussian approximation.

$E_\mathrm{out}$ is determined from the time of flight between the TOF sensor planes, $t_2 - t_1$, as 

\begin{equation}
E_\mathrm{out} = \frac{m_p c^2}{\sqrt{1 - (v/c)^2}} - m_p c^2 \quad \mathrm{with} \quad v = \frac{L}{t_2-t_1}, 
\end{equation}

\noindent where $L$ is the distance between the TOF sensor planes and $m_p=938$\,MeV/c$^2$ is the proton  mass. 
Energy loss in 1\,mm of silicon or 1\,m of air is on the order of 0.5\,MeV for protons with an energy between 100\,MeV and 200\,MeV \parencite{pstar}. 
We therefore neglect energy loss in the trackers, the TOF sensors, and the surrounding air. By first order error propagation one has

\begin{equation}\label{eq:sigmatof}
\sigma_{E_\mathrm{out},\mathrm{TOF}}(E_\mathrm{out}) = \left( \left| \frac{\rmd E}{\rmd t_1}  \right|^2 \sigma^2_{t_1} + \left| \frac{\rmd E}{\rmd t_2}  \right|^2 \sigma^2_{t_2} \right)^{1/2} = \frac{1}{m^2_p c^3} (E_\mathrm{out}^2 + 2m_p c^2 E_\mathrm{out})^{3/2} \; 	\frac{\sigma_t}{L}, 
\end{equation}

\noindent where $\sigma_t^2=\sigma_{t_1}^2 + \sigma_{t_2}^2$ is the TOF measurement error stemming from the time resolution of the two sensors. 
Note that the  error on $E_\mathrm{out}$ directly scales with $L$ so that a greater spacing between the  TOF planes will reduce the energy measurement error. Throughout this work, 
we report the velocity error $\sigma_t/L$ in units of ps/m, rather than $\sigma_t$ alone.  

Energy straggling of protons inside the imaged object and detectors leads to a variation of $E_\mathrm{out}$ and acts as another source of statistical error in the data \parencite{Schulte2005}. The energy spread due to straggling can be modelled theoretically via partial differential equations \parencite{Tschalar1968} and solved analytically \parencite{Payne1969}, yielding

\begin{equation}\label{eq:stragglingintegral}
\sigma_{E_\mathrm{out}, \mathrm{straggling}}(E_\mathrm{out}) = \chi_1^2(E_\mathrm{out}) \int_{E_\mathrm{out}}^{E_\mathrm{in}} \frac{\chi_2(E)}{\chi_1^3(E)} \rmd E,  
\end{equation}

\noindent where $\chi_1(E)$ and $\chi_2(E)$ are defined as follows:

\begin{align}
\chi_1(E) &= K \frac{1}{\beta^2} \left[ \ln \left( \frac{2m_p c^2\beta^2}{I (1-\beta^2)} - \beta^2 \right) \right] \\
\chi_2(E) &= K \frac{1 - \beta^2/2}{1 - \beta^2} \quad \mathrm{with} \quad \beta=\frac{v}{c} = \left[ 1 - \left(\frac{m_p c^2}{m_p c^2 + E}\right)^2 \right]^{1/2}. 
\end{align}

\noindent Here, $I$ is the ionisation potential of the target material of interest which we approximate to be water ($I$=75\,eV, \textcite{Berger1993}), $m_p$ is the proton mass, $\beta$ the proton velocity relative to the speed of light, and $K = 170$\,MeV/cm is a constant. Note that $\chi_1(E)$ is identical to the stopping power given by the Bethe-Bloch equation \parencite{Paganetti2012book}. We solved equation~\ref{eq:stragglingintegral} numerically. 

Because the protons' energy loss, $\Delta E = E_\mathrm{out} - E_\mathrm{in}$, is obtained as the difference between the measured exit energy and the incident energy determined from the beam energy provided by the proton accelerator, the energy spread of the incident beam (standard deviation $\sigma_\mathrm{beam}$; hereafter denoted as beam spread) acts as an additional error. 
As a figure of merit, the relative energy spread of a therapeutic proton beam, $\delta E_\mathrm{beam} = \sigma_\mathrm{beam}(E_\mathrm{in}) / E_\mathrm{in}$ is typically 0.5\% to 1\% of the beam energy \parencite{Schippers2018a}.  
The overall error in the energy loss is the squared sum of the contributions due to straggling, TOF measurement error, and beam spread, i.e.

\begin{equation}\label{eq:sigmaEtotal}
\sigma^2_{\Delta E}(E_\mathrm{out}) = \sigma^2_{E_\mathrm{out}, \mathrm{straggling}}(E_\mathrm{out})  + \sigma^2_{E_\mathrm{out},\mathrm{TOF}}(E_\mathrm{out}) + (\delta E_\mathrm{beam} E_\mathrm{in})^2. 
\end{equation}

\noindent Note that we disregard the noise induced by multiple Coulomb scattering on target nuclei in the scanned object \parencite{Dickmann2019} for simplicity and because it depends on the shape and composition of the object. Therefore, it is not purely characteristic of the TOF proton CT scanner. 

In distance driven proton CT reconstruction \parencite{Rit2013}, the WEPL value in a pixel is obtained by converting energy loss to WEPL and averaging the WEPL of all protons binned in that pixel. The relation between WEPL and exit energy $E_\mathrm{out}$ is analytically given by 

\begin{equation}\label{eq:wepl_E}
\mathrm{WEPL}(E_\mathrm{out}) = \int_{E_\mathrm{in}}^{E_\mathrm{out}} 1/S_w(E) \rmd E.
\end{equation}

\noindent In this work, we performed the integration in equation~\ref{eq:wepl_E} numerically using the stopping power $S_w(E)$ of water from NIST's PSTAR table \parencite{pstar}. We obtained the conversion from WEPL to $E_\mathrm{out}$ by numerical inversion. 
The energy error propagates (first order) to a WEPL error, $\sigma_\mathrm{WEPL}$, as 

\begin{equation}\label{eq:sigmawepl}
\sigma^2_\mathrm{WEPL}(E_\mathrm{out}) = \frac{\sigma^2_{\Delta E}(E_\mathrm{out})}{S_w^2(E_\mathrm{out}) N} = \frac{\sigma^2_{\Delta E}(E_\mathrm{out})}{S_w^2(E_\mathrm{out}) \Phi \Delta\xi^2}, 
\end{equation}

\noindent where $S_w(E_\mathrm{out})$ is the proton stopping power in water at the energy $E_\mathrm{out}$.  The number of protons $N$ contributing to a given pixel has been re-expressed as the product of fluence $\Phi$ and area of reference, in this case the pixel area $\Delta\xi^2$. 

We also consider a proton CT system using a calorimeter as energy detector, e.g. a scintillator coupled with photo multipliers \parencite{Johnson2016}.  \textcite{Bashkirov2016} parametrise the measurement uncertainty via a constant relative error,  $\delta E_\mathrm{out,cal} \equiv \sigma_{E_\mathrm{out,cal}}/E_\mathrm{out}$ on the order of a few percent. The $\sigma^2_{E_\mathrm{out},\mathrm{TOF}}$ term in equation~\ref{eq:sigmaEtotal} is replaced by $(\delta E_\mathrm{out,cal} E_\mathrm{out})^2$ in this case.

\subsection{Noise reconstructions}\label{sec:recon}
In statistical terms, the noise in the reconstructed proton CT images, i.e. RSP precision, is the variance of the reconstructed RSP value in each image pixel. 
This can be obtained, e.g., by simulating many independent proton CT acquisitions and calculating the voxel-wise root-mean-square-error (RMSE) over the ensemble of reconstructed images. However, it is also possible to reconstruct noise images \parencite{Wunderlich2008} starting from noise projections. \textcite{Radler2018} have implemented this method for proton CT. 
We implemented the noise reconstruction using a cone beam geometry and a full circular ($2\pi$) acquisition trajectory with radius $R_0$, but we only reconstructed the central slice. For the purpose of this work, we therefore limit our geometry to a two-dimensional description. We briefly describe our noise reconstruction in the following and  refer to \textcite{Wunderlich2008,Radler2018} for a detailed description of the underlying considerations. 

\begin{figure}
\centering
\includegraphics[width=0.4\textwidth]{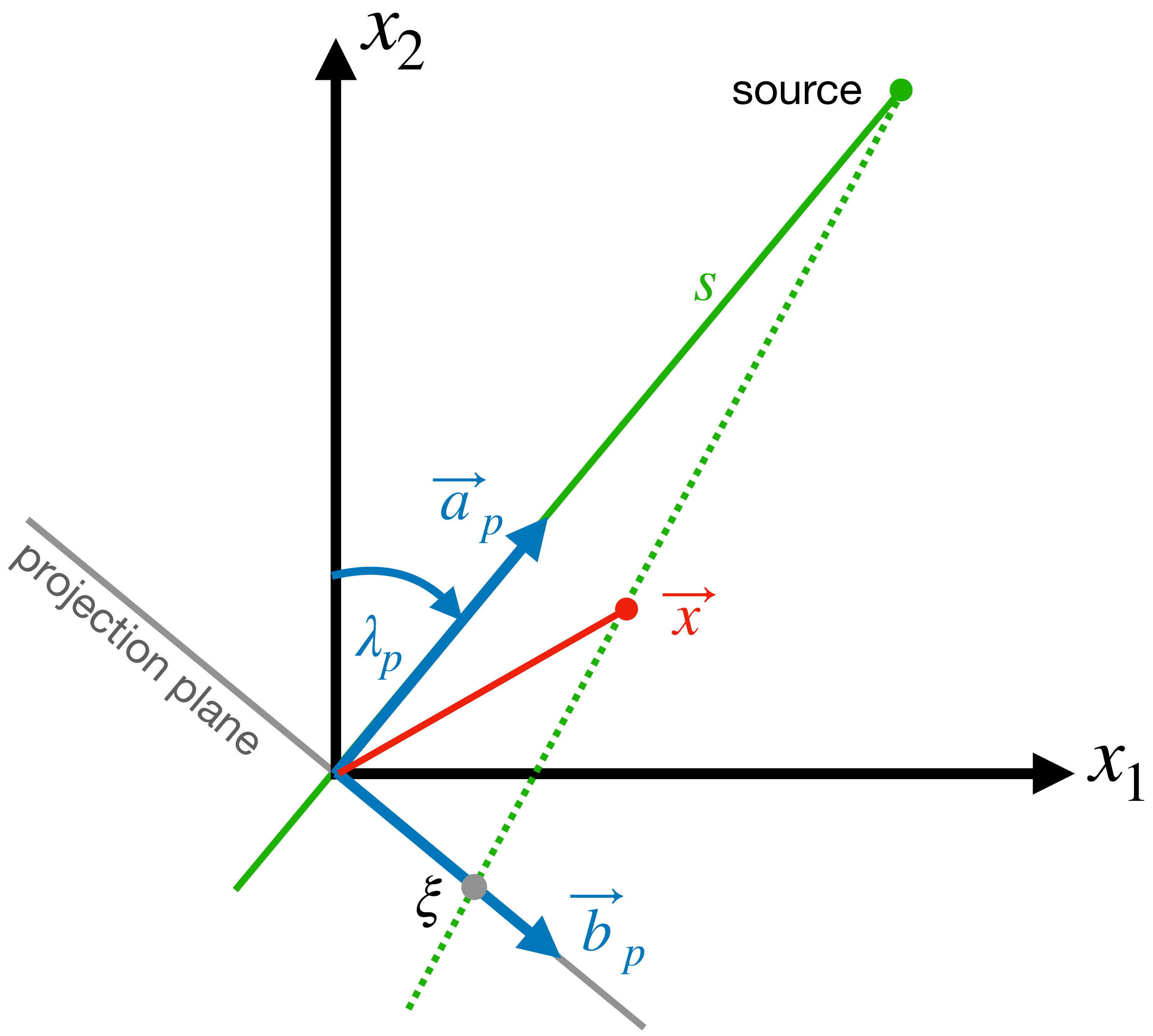}
\caption{Scanner geometry considered in the noise reconstructions. }
\label{fig:geometry}
\end{figure}

An illustration of the geometry is shown in figure~\ref{fig:geometry}. 
We place the rotation axis at the origin $(0,0)$. The source angle is denoted with $\lambda_p$ and increases clockwise, $s$ is the source to centre distance, and $\vec{a}_{\lambda_p}$ is the source position in two dimensional Cartesian coordinates, i.e. $\vec{a}_{\lambda_p} = s (\sin \lambda_p, \cos \lambda_p)^T$.  The basis vector of the projection plane (1D) going through $(0,0)$ is $\vec{b}_{\lambda_p}=(\cos \lambda_p, -\sin \lambda_p)^T$ and $\xi$ refers to the coordinate in the projection frame. For a given point $\vec{x}=(x_1,x_2)^T$ and source angle $\lambda_p$, the coordinate $\xi$ in the projection frame is given by 

\begin{equation}
\xi = \vec{x}\cdot \vec{b}_{\lambda_p}  \frac{s}{s - \vec{x}\cdot \vec{a}_{\lambda_p}} = \frac{s \left( x_1 \cos \lambda_p - x_2 \sin \lambda_p \right)}{s - x_1\sin\lambda_p - x_2\cos\lambda_p}. 
\end{equation}

\noindent The position $\xi$ and angle $\lambda_p$ are sampled uniformly, i.e. 

\begin{align}
\lambda_p &= (p-1) \Delta\lambda, \quad p=1,2,\dots,P\\
\xi_j &= (j + 1/2) \Delta\xi, \quad j=-J, -J+1, \dots, J-1 
\end{align}

\noindent with $\Delta\lambda$ the angular spacing between projections, $P$ the number of projections, $\Delta\xi$ the pixel width in a projection, and $2J$ the number of sampled positions in the central projection plane.

We assumed that projections are filtered with an apodized ramp filter \parencite{Ramachandran1971}, 

\begin{equation}
h_F(\xi_j) = h_F((j+1/2) \Delta\xi) = 
\begin{cases}
1/(2\Delta\xi)^2  &\mathrm{for}\, j=0, \\
0 &\mathrm{for}\,j\, \text{even and}\, j\neq0 \\
-1 / (j\pi\Delta\xi)^2 &\mathrm{for}\, j\, \mathrm{odd},
\end{cases}
\end{equation}

\noindent and that images are reconstructed with the Feldkamp-Davis-Kress (FDK) algorithm \parencite{Feldkamp1984}, which is the case in the distance driven binning algorithm in \parencite{Rit2013}. For simplicity, we treated proton trajectories as straight lines and therefore only used one projection image per projection angle corresponding to the central depth, unlike \textcite{Radler2018} who considered the influence of multiple Coulomb scattering. 

In general, the variance of the RSP value at point $\vec{x}$ is calculated by backprojecting the variance and covariance of the filtered projections, where the covariance arises from the combination of interpolation and filtering. We use a simplified version of the noise reconstruction which replaces the covariance term (for bilinear interpolation of projections filtered by the same apodized ramp filter) by an effective factor $f_\mathrm{interp}=2/3 - 2/\pi^2$ \parencite{Radler2018}. This avoids interference patterns in the noise distributions due to correlations among pixels and makes the results easier to interpret. In particular, the RSP variance at  $\vec{x}$ is

\begin{equation}\label{eq:etabackproj}
\mathrm{Var}_\mathrm{RSP}(\vec{x}) = f_\mathrm{interp} \frac{\Delta \lambda^2}{4} 
\sum_{p=1}^P \left( \frac{\|\vec{a}_{\lambda_p} \|}{\|\vec{x} - \vec{a}_{\lambda_p} \|}  \right)^4 V_p(\xi_k). 
\end{equation}

\noindent The term $V_p(\xi_k)$ is the variance in the filtered projections at $\xi_k$,

\begin{equation}\label{eq:Vp}
V_p(\xi_k) = (\Delta\xi)^2 \sum_{j=-J}^{J-1} h_F^2(\xi_k-\xi_j) \frac{\|\vec{a}_{\lambda_p}\|^2}{\|\vec{a}_{\lambda_p} \|^2 + \xi_j^2}\mathrm{Var}(\lambda_n, \xi_j), 
\end{equation}

\noindent where $\mathrm{Var}(\lambda_n, \xi_j)$ refers to the noise projection containing the WEPL variance, $\frac{\|\vec{a}_{\lambda_p}\|^2}{\|\vec{a}_{\lambda_p} \|^2 + \xi_j^2}$ is a weighting factor \parencite{Kak2001}.

\subsection{Simulations and phantoms}\label{sec:simuphan}

We used a cylindric water phantom with a 20\,cm diameter and with infinite axial extension.
To simulate noise projections, we forward projected RSP values of the phantom to obtain a projection image containing WEPL values, converted these into energy loss via the inverse of equation~\ref{eq:wepl_E}, and calculated the variance projections $\mathrm{Var}(\lambda_n, \xi_j)$ via equation~\ref{eq:sigmawepl} as described in section~\ref{sec:tofstraggling}. 
We approximated proton paths as straight lines thereby neglecting the effect of multiple Coulomb scattering. 
This simplified the reconstruction because distance driven binning was not necessary. 
Energy loss and scattering in the air was neglected, given that 1\,m of air corresponds to only 1\,mm of water in terms of mass density.  
We used 360 projection angles uniformly distributed over a full circular trajectory and assumed a source-to-center distance of 200\,cm, which is representative of many modern therapeutic proton accelerators. 
The pixel size was $1\times1$\,mm$^2$, both in the  projection images and in the reconstructed images. 
We considered an energy spread of the proton beam of $\delta E_\mathrm{beam}=0.5$\% as a figure of merit \parencite{Schippers2018a}.  

We determined the number of protons contributing to a pixel in equation~\ref{eq:sigmawepl} in relation to the dose at the phantom centre of 10\,mGy accumulated over an entire proton CT scan. 
Specifically, the fluence at the exit of the cylinder, $\Phi_{\mathrm{out}}$, i.e. the fluence of protons measurable by the energy detector, depends on the thickness of material traversed by the protons expressed in WEPL, which we denote with $W(\xi)$ and which depends on the location $\xi$ in the projection plane (but not on the source position $\lambda$ because the scanned object is a cylinder). Assuming a spatially constant initial fluence, $\Phi_\mathrm{in}$, one can relate $\Phi_{\mathrm{out}}(W(\xi))$ and the proton fluence at the cylinder centre $\Phi_\mathrm{centre}$: 

\begin{equation}\label{eq:attenuation}
\Phi_{\mathrm{out}}(W(\xi)) = \Phi_\mathrm{in} \exp(-\kappa W(\xi)) = \Phi_\mathrm{centre} \exp(-\kappa (W(\xi)-R)), 
\end{equation}

\noindent where  $R$ is the cylinder radius and $\kappa=0.01$\,cm$^{-1}$ is a coefficient describing attenuation due to nuclear interactions in water \parencite{Schulte2005,Quinones2016}. 
$\Phi_\mathrm{centre}$ is related to the dose $D_\mathrm{centre}$ accumulated at the cylinder centre during a full pCT scan ($P$ projections) \parencite{Schulte2005}: 

\begin{equation}\label{eqn:centraldose}
\Phi_\mathrm{centre} =  \frac{D_\mathrm{centre}\rho}{P\left[ S(E_\mathrm{centre})+\kappa\gamma E_\mathrm{centre}\right]},
\end{equation}

\noindent where $S(E_\mathrm{centre})$ is the proton stopping power for the  proton energy at the centre, $\rho$ is the mass density of the cylinder material (in our case water, i.e. $\rho=1$\,g/cm$^3$), $\gamma$=0.65 is the fraction of energy transferred to secondary particles during nuclear interactions (in the energy range up to 200\,MeV, in water-like material). In a water cylinder with $R=10$\,cm radius and for a 200\,MeV proton beam, one has $E_\mathrm{centre}\approx 151$\,MeV and consequently a proton stopping power $S(E_\mathrm{centre})$ in water of about 5.4\,MeV\,cm$^2$/g \parencite{pstar}. 
With these numbers, equation~\ref{eqn:centraldose} yields a fluence at the centre of 270\,protons per mm$^2$ per projection, which corresponds well to the numbers in \textcite{Schulte2005} estimated via Geant4 Monte Carlo simulation \parencite{Agostinelli2003}.

The full expression for the modelled variance projection is obtained by combining equations~\ref{eq:sigmawepl}, \ref{eq:attenuation}, and~\ref{eqn:centraldose},

\begin{equation}\label{eq:varg}
\mathrm{Var}(\lambda_p, \xi_j) = \frac{\sigma^2_{E_\mathrm{out}}(\lambda_p, \xi_j)}{S_w^2(E_\mathrm{out}(\lambda_p, \xi_j))} \frac{P \left[ S(E_\mathrm{centre})+\kappa\gamma E_\mathrm{centre} \right]}{D_\mathrm{centre} \rho  \Delta\xi^2} \exp(\kappa \left(W(\lambda_p, \xi_j)-R) \right),
\end{equation} 

\noindent where $W(\lambda_p, \xi_j)$ is the WEPL at sampling point $\xi_j$ in projection $\lambda_p$ and $E_\mathrm{out}(\lambda_p, \xi_j)$ is obtained from $W(\lambda_p, \xi_j)$ via the inverse of equation~\ref{eq:wepl_E}.

\section{Results}

\begin{figure}
\centering
\includegraphics[width=0.48\textwidth]{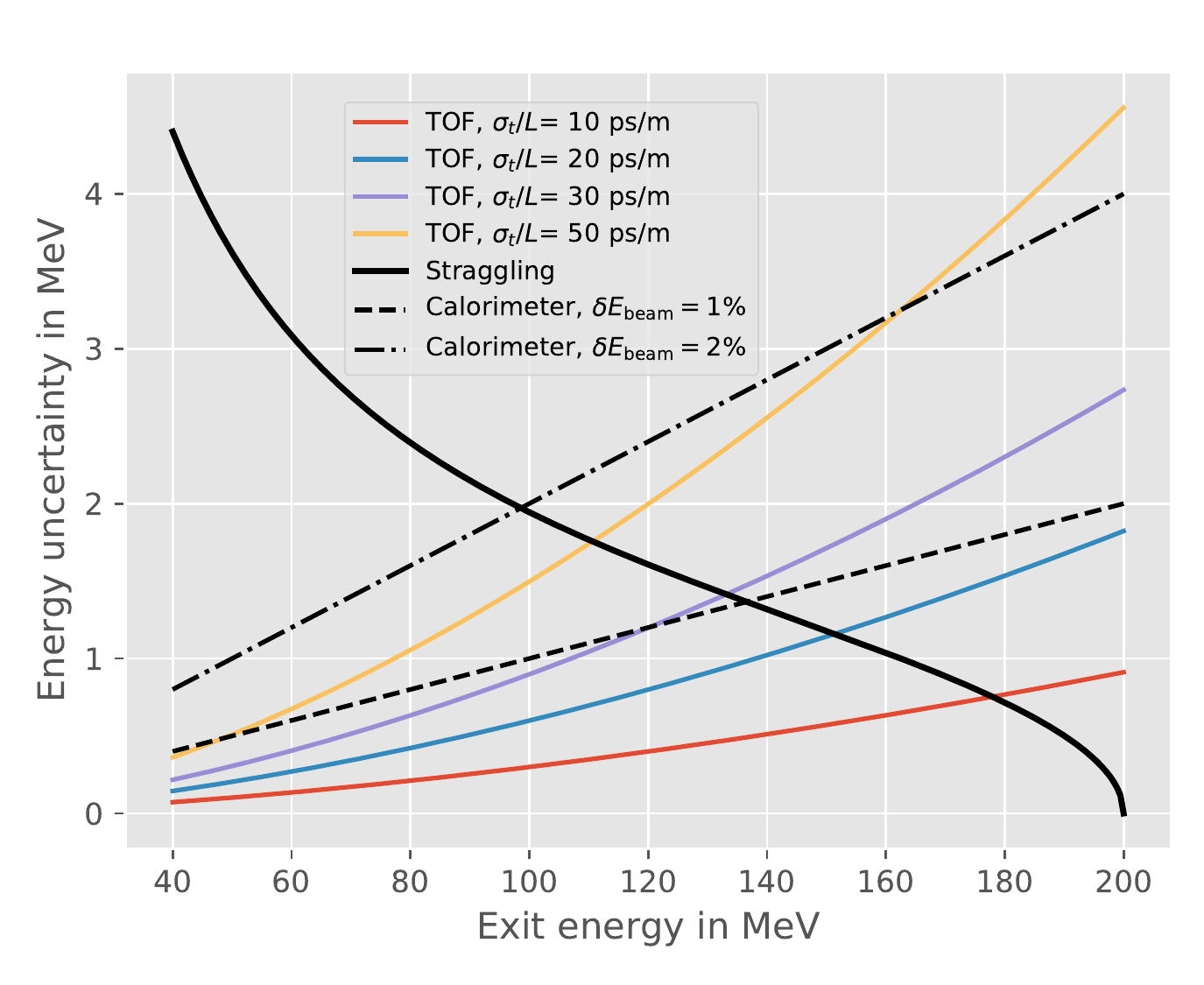}
\hfill
\includegraphics[width=0.48\textwidth]{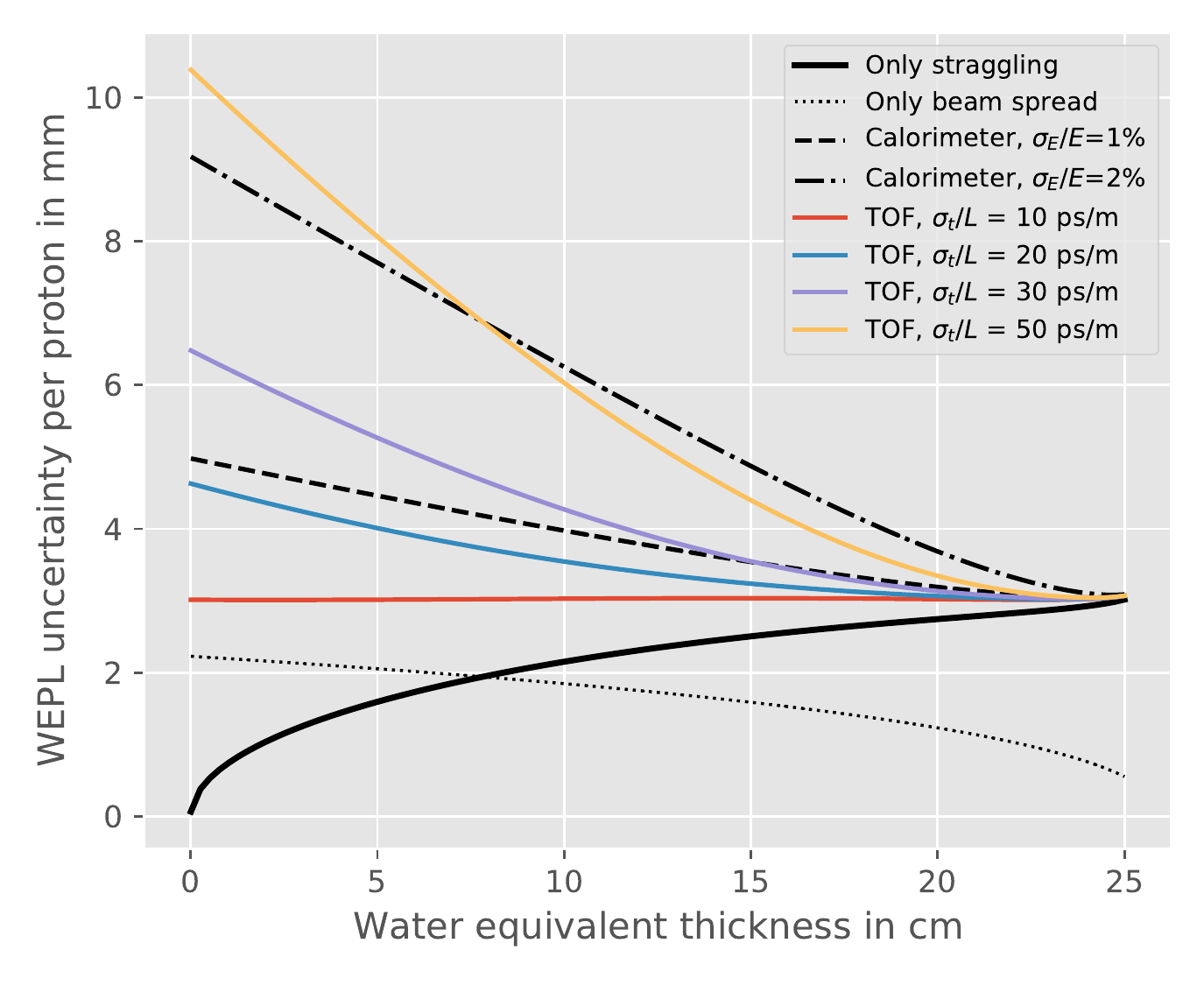} \\
\caption{Left: Energy uncertainty (1 sigma) due to straggling and energy measurement error for different velocity errors $\sigma_t/L$ and a beam energy of 200\,MeV. The dashed and dash-dotted lines depict a 1\% and 2\% uncertainty of a calorimeter detector, respectively. Right: WEPL uncertainty (1 sigma) of a proton CT system (equation~\ref{eq:sigmawepl}) as a function of WEPL for a beam energy of 200\,MeV and different velocity errors  $\sigma_t/L$. Right: WEPL uncertainties for a calorimeter-based system (dashed, dash-dotted) are shown for comparison. }
\label{fig:tof_physics_energy}
\end{figure}

The left panel in figure~\ref{fig:tof_physics_energy} shows the energy uncertainty due to the energy measurement error, either via TOF (solid) or via calorimeter (dashed, dash-dotted),  and due to straggling (solid black) for an incident beam energy of 200\,MeV. Throughout the results section, we report all uncertainties as one sigma errors, i.e. one standard deviation. 
It is intuitively clear that the energy uncertainty due to TOF error decreases with decreasing exit energy because the lower the proton's energy, the longer its time-of-flight so that the time resolution of the TOF sensors has less impact on the estimated velocity. 
The limit is that protons still need to reliably traverse the scanned object rather than lose all their energy inside. The lowest depicted energy value of 40\,MeV corresponds to a residual range of about 1.5\,cm. 
Straggling, on the other hand, increases with the degree to which a proton has been slowed down due to electromagnetic interactions and thus increases with decreasing exit energy. 

The right panel in figure~\ref{fig:tof_physics_energy} shows how the combined energy uncertainty (due to straggling, measurement error, and beam spread) translates into WEPL uncertainty as a function of the phantom's water equivalent thickness, again for a beam energy of 200\,MeV. 
In thin objects, it is mainly the TOF error which dominates the WEPL uncertainty while it is mainly straggling in thicker objects. For velocity errors larger than 50\,ps/m, the measurement error is more dominant than straggling for almost all WEPLs. Overall, the WEPL precision of a 20-30\,ps/m TOF proton CT system is comparable to a calorimeter-based system with 1\% energy measurement error, while a 2\%  calorimeter error corresponds to a velocity error $\sigma_t/L$ of about 50\,ps/m.   

\begin{figure}
\centering
\includegraphics[width=0.32\textwidth]{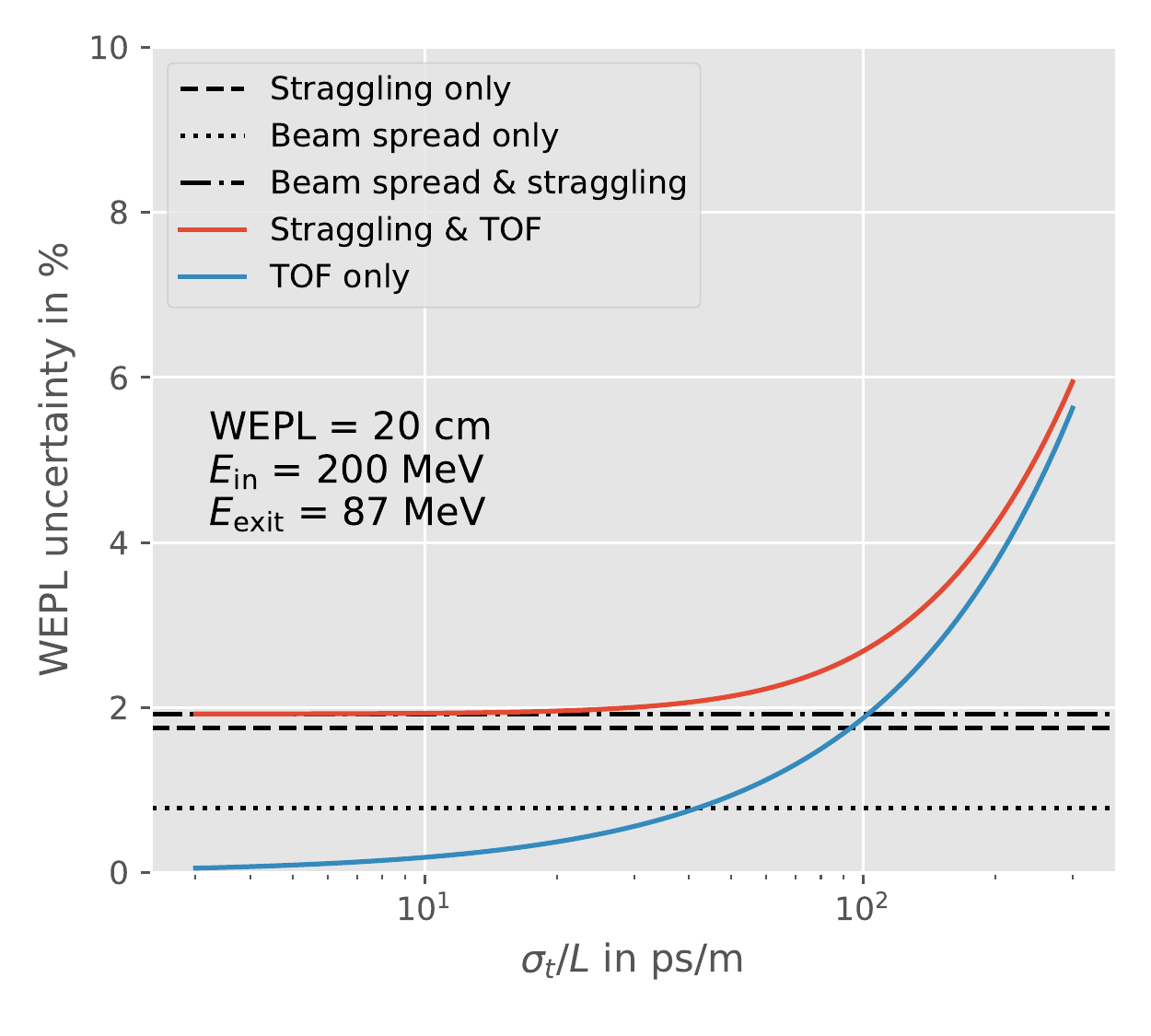}
\hfill
\includegraphics[width=0.32\textwidth]{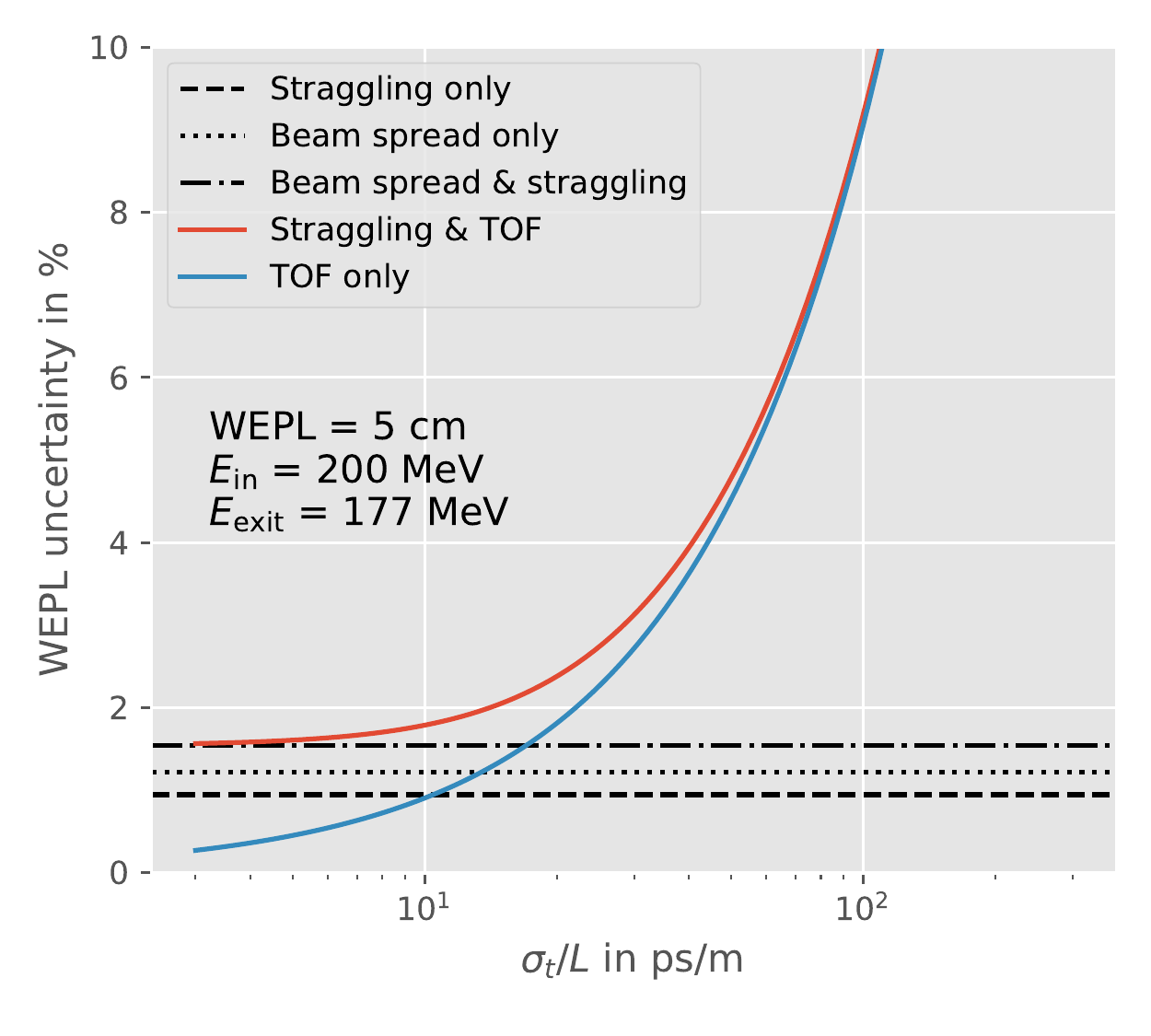}
\hfill
\includegraphics[width=0.32\textwidth]{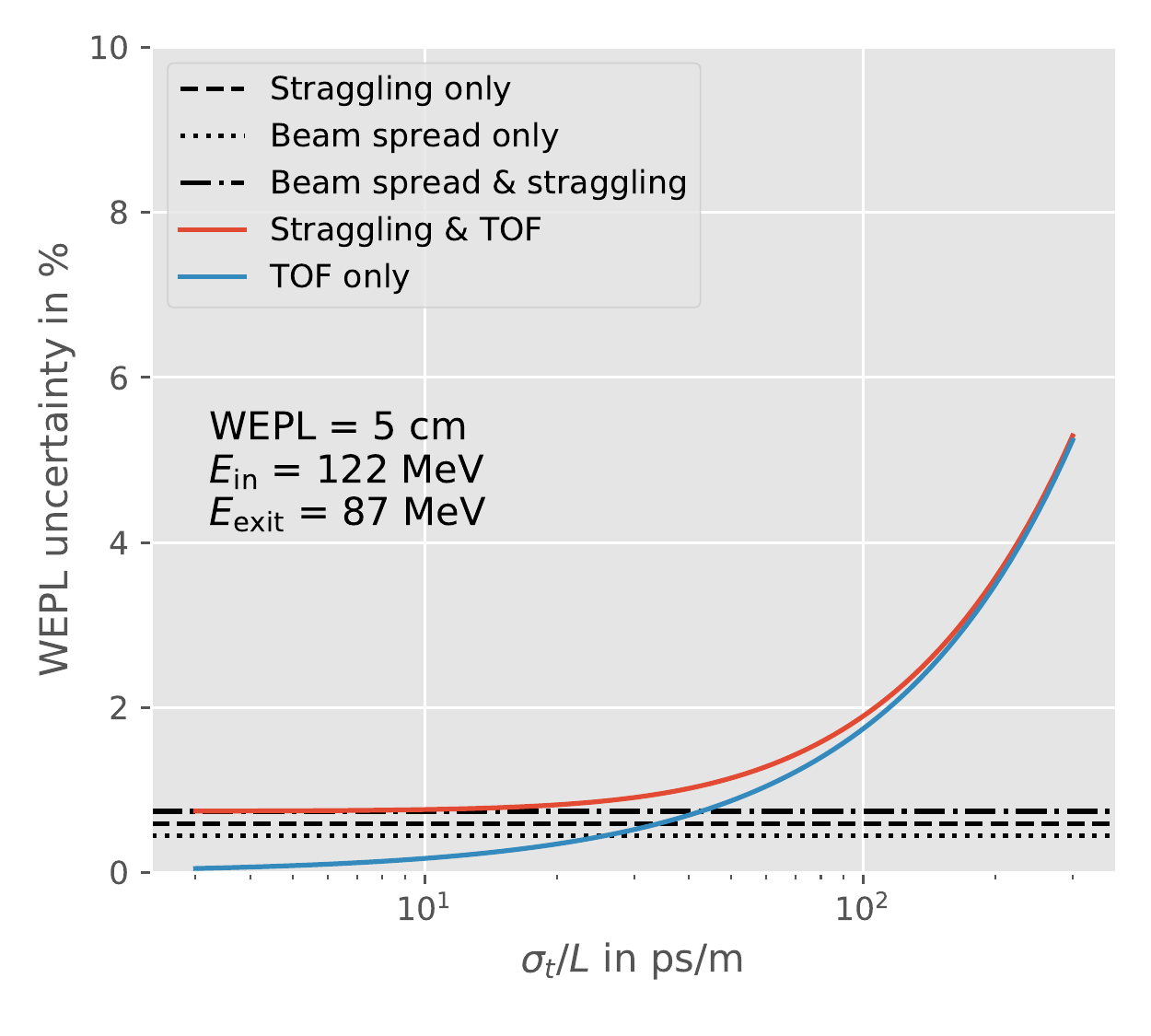}
\caption{WEPL uncertainty in a TOF proton CT as a function of velocity error $\sigma_t/L$ for three different combinations of object thickness and beam energy. }
\label{fig:wepluncert_sigmatof}
\end{figure}

Figure~\ref{fig:wepluncert_sigmatof} compares the three contributions to WEPL uncertainty, i.e. straggling, TOF measurement error, and beam spread, as a function of the velocity error $\sigma_t/L$ for an  object thickness of 5\,cm and 20\,cm, as indicated in the plots. 
With a WEPL of 20\,cm, the point of equal contribution (straggling + beam spread = TOF error) is found at $\sigma_t/L=$100\,ps/m. With a WEPL of  5\,cm, on the other hand, the TOF error begins to dominate already for $\sigma_t/L\approx20$\,ps/m. Lowering the beam energy to 122\,MeV, which corresponds to the same exit energy of 86\,MeV as for the 20\,cm/200\,MeV case, shifts the equal contribution point to about 40\,ps/m. 

\begin{figure}
\centering
\includegraphics[width=0.48\textwidth]{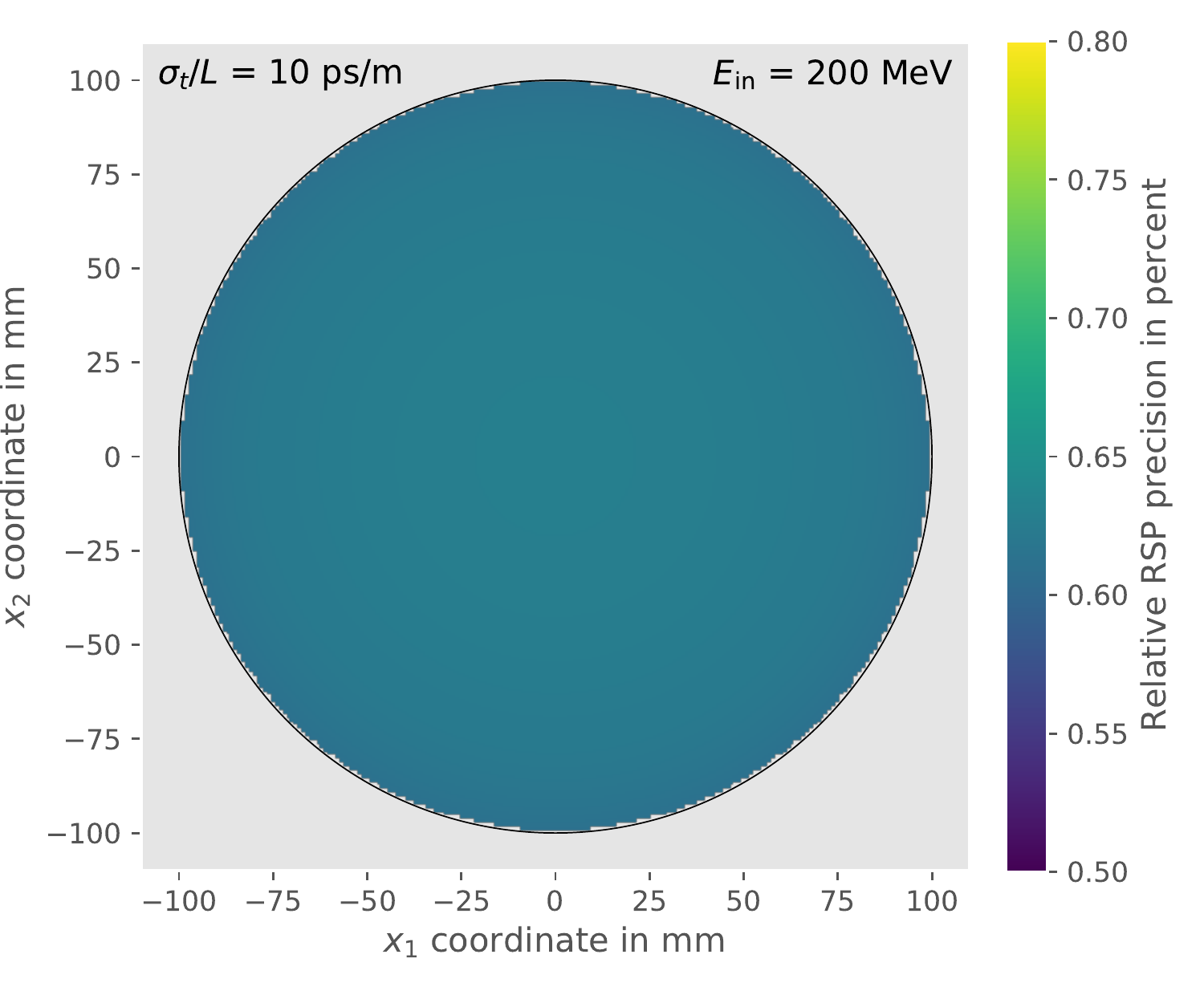}
\hfill
\includegraphics[width=0.48\textwidth]{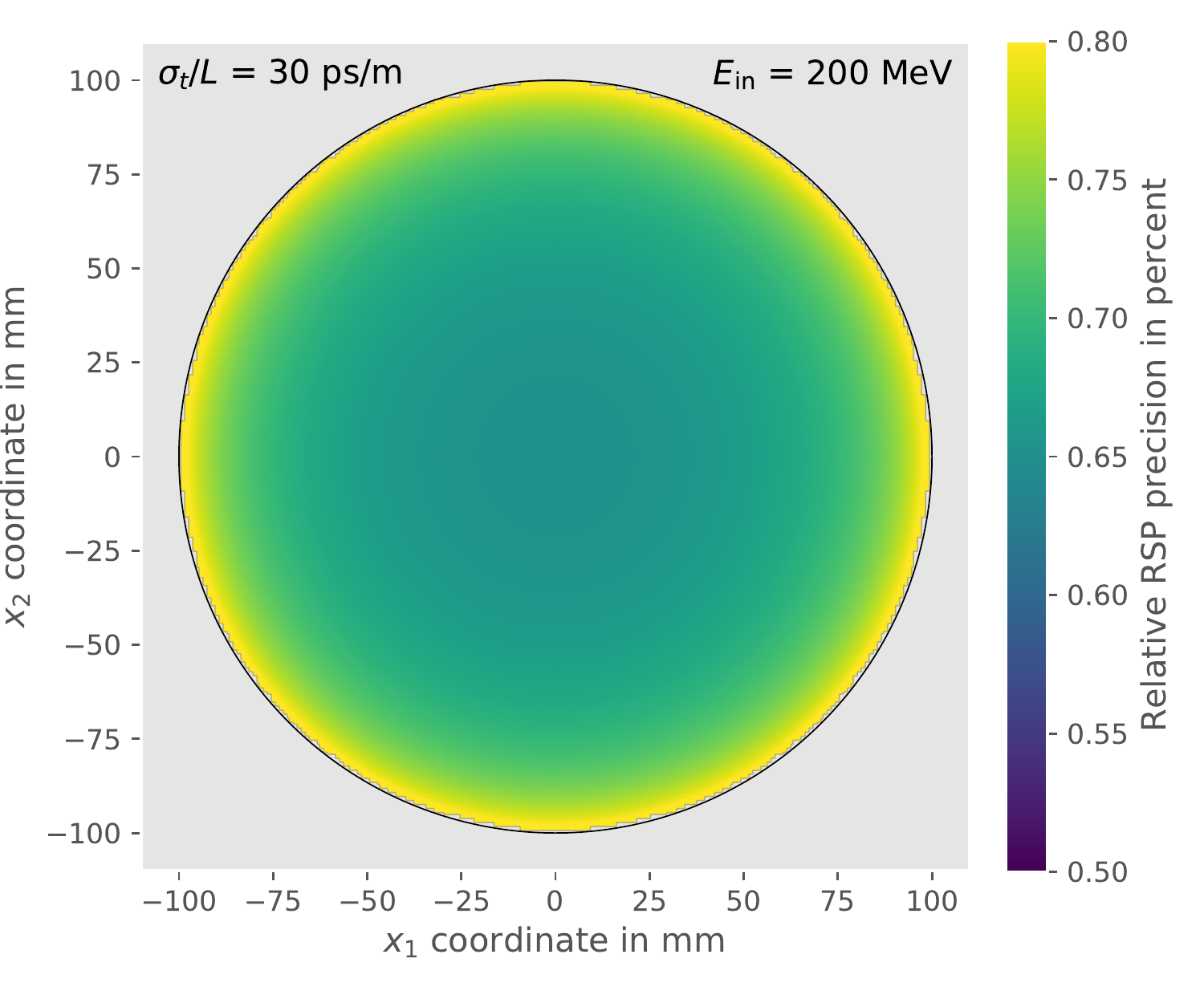}
\\
\includegraphics[width=0.48\textwidth]{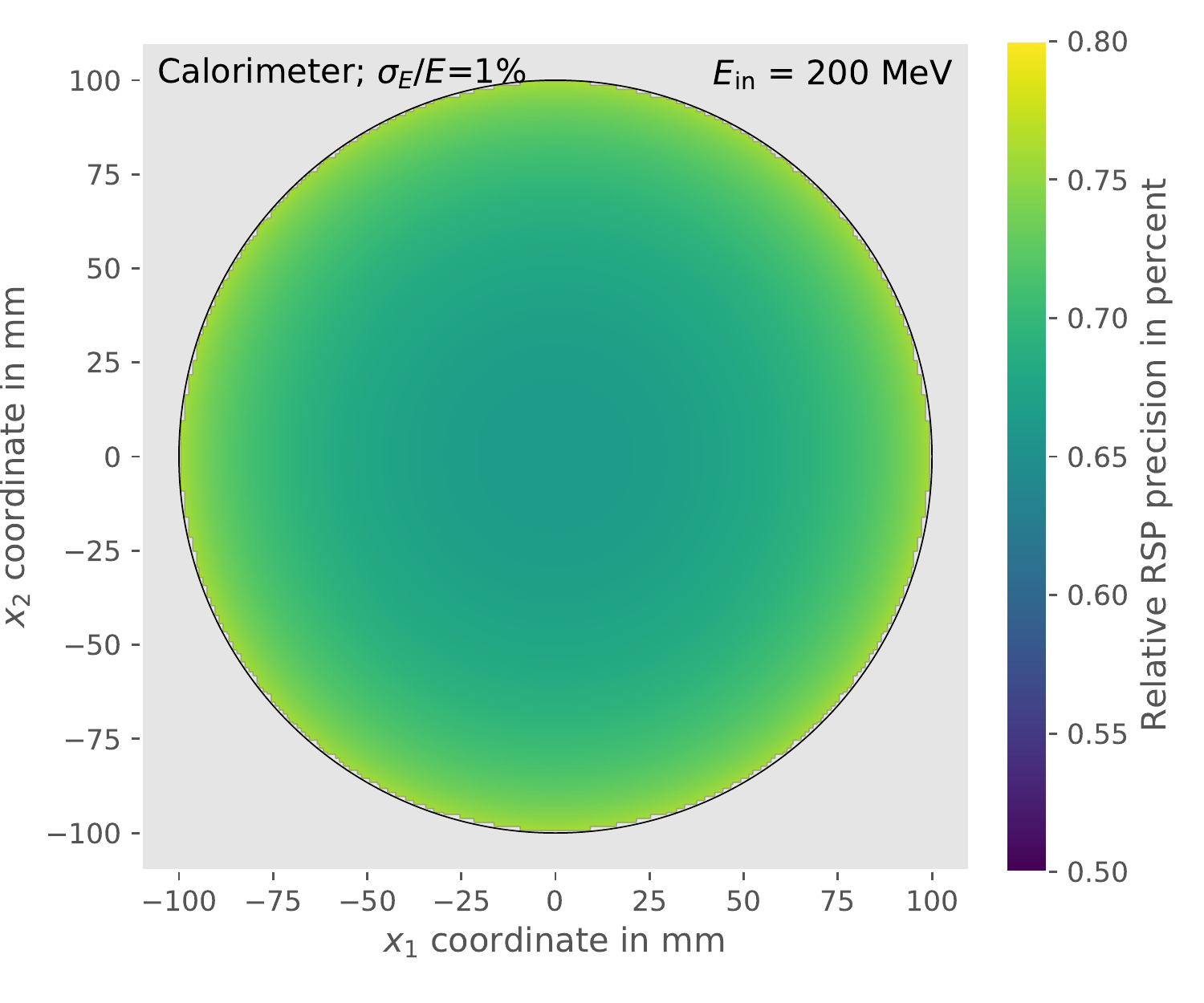}
\hfill
\includegraphics[width=0.48\textwidth]{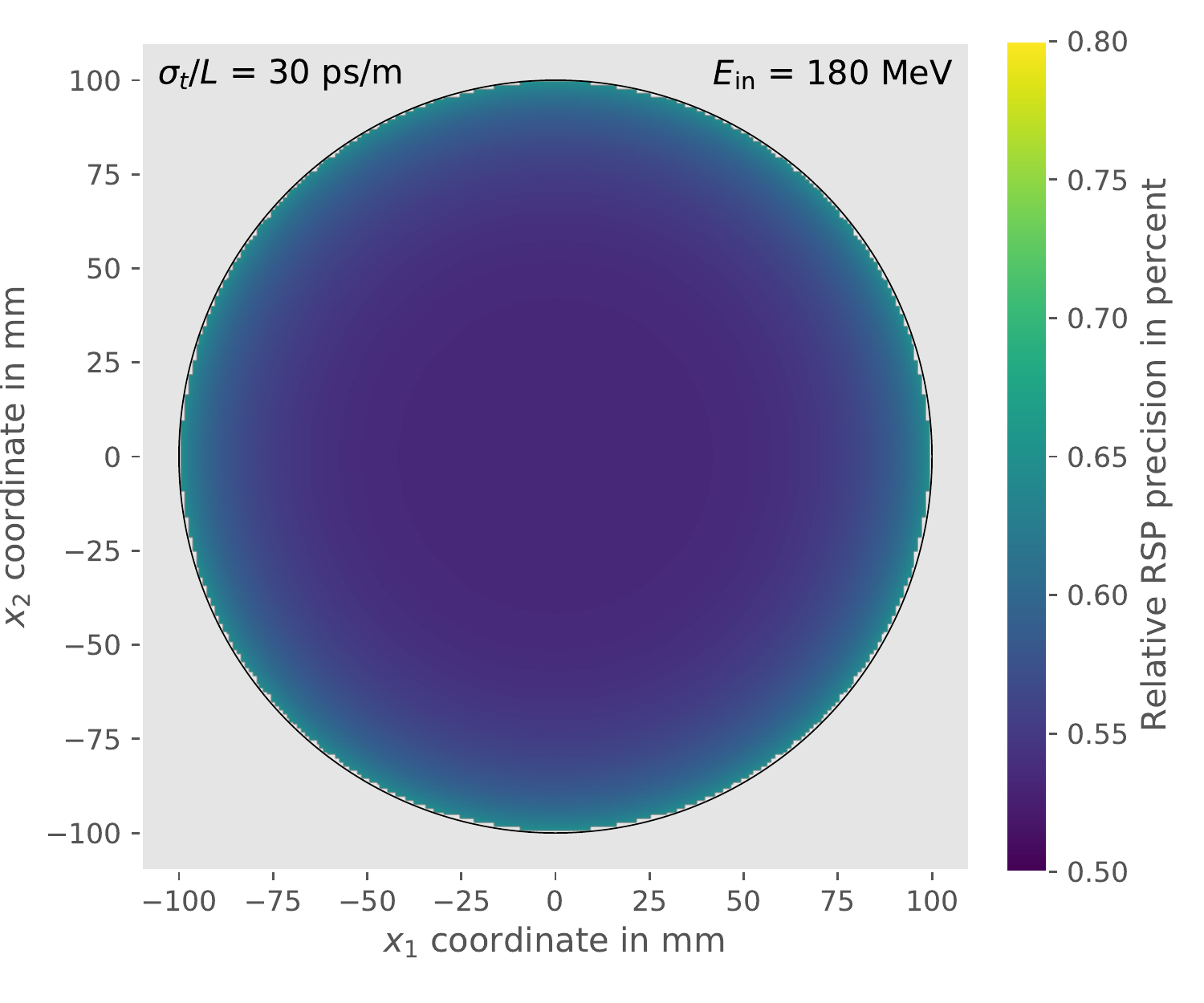}\\
\includegraphics[width=0.48\textwidth]{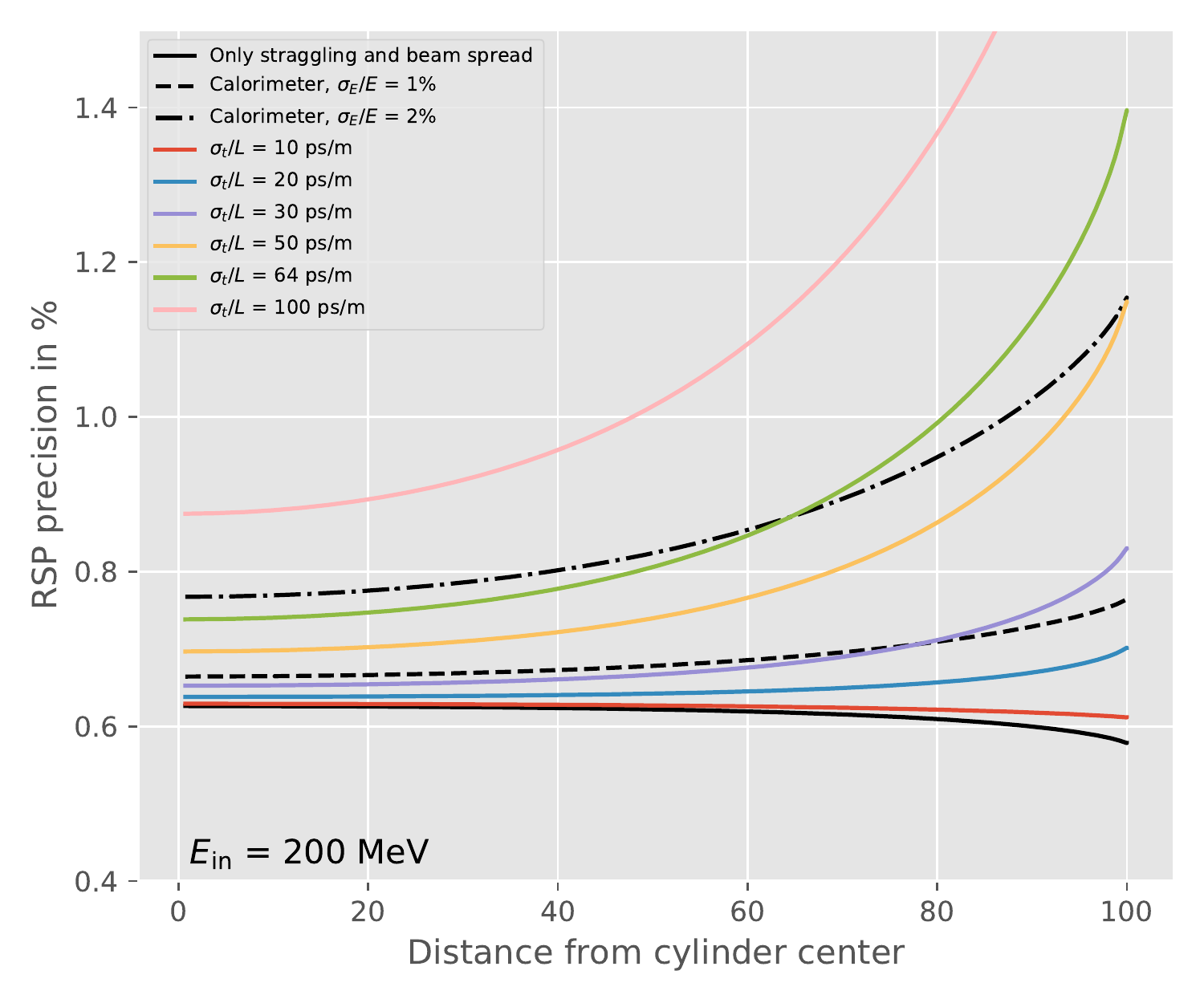}
\hfill
\includegraphics[width=0.48\textwidth]
{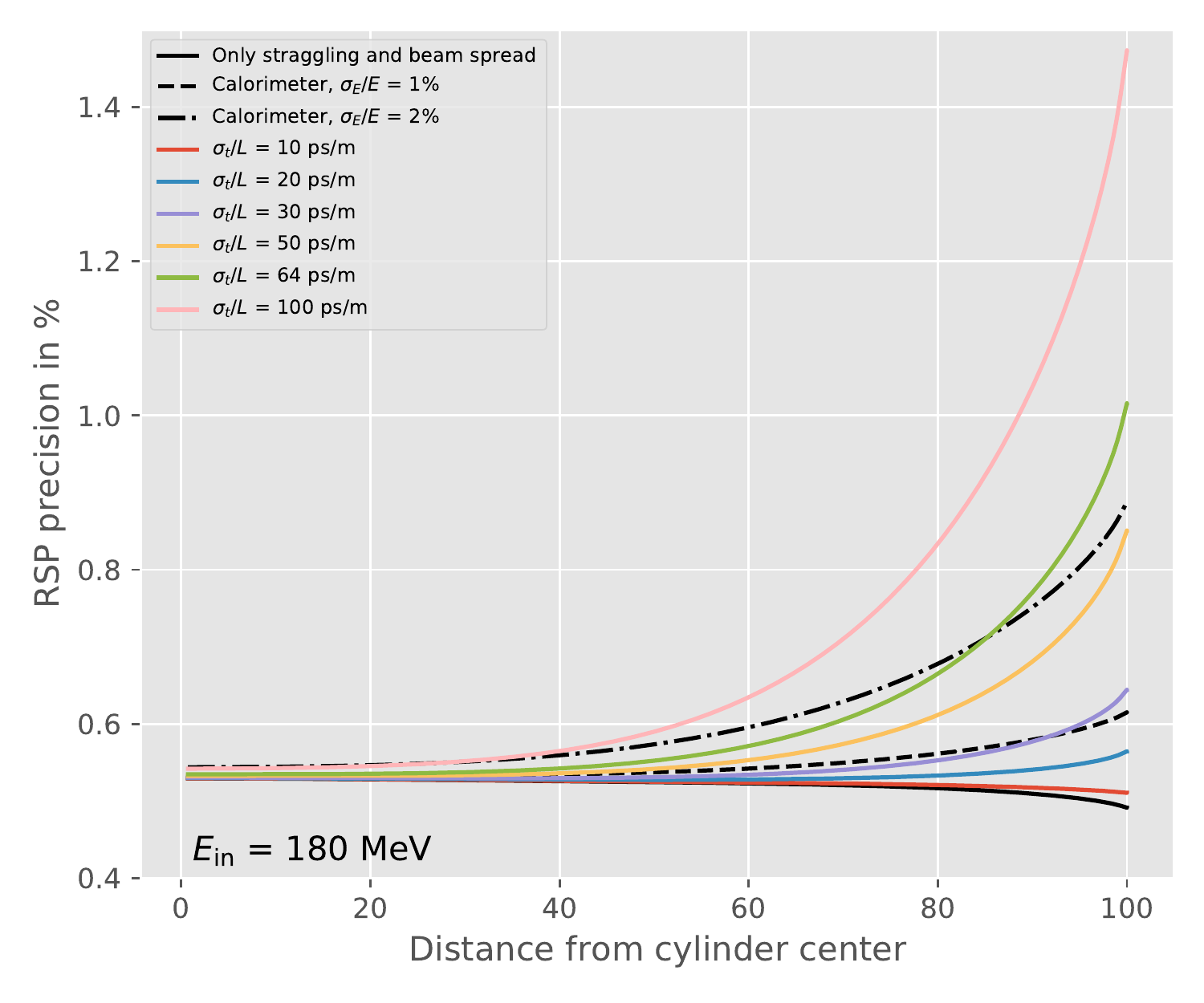}
\caption{Upper panels: noise reconstructions of a water cylinder of 20\,cm diameter for different velocity errors  $\sigma_t/L$ and calorimeter measurement errors as well as different incident beam energies as indicated in the graphics. Lower panels: Radial profiles of the RSP precision for 200\, MeV (left) and 180\,MeV (right) beam energy. }
\label{fig:noiserecon}
\end{figure}

The four upper panels in figure~\ref{fig:noiserecon} show the RSP precision maps for a water cylinder with 20\,cm diameter and different velocity errors  $\sigma_t/L$ (TOF system) and $\delta E=\sigma_E/E$ (calorimeter system), as indicated in the graphics. 
The noise images were simulated as described in section~\ref{sec:simuphan} and reconstructed according to  section~\ref{sec:recon}. The two lower panels show radial noise profiles for different values of $\sigma_t/L$, for calorimeter precisions of $\delta E=1$\% and $\delta E=2$\%, and the idealised case where only straggling and the beam's energy spread contribute to image noise.  
Noise is distributed quite evenly for $\sigma_t/L$ between 10~and 20\,ps/m, while there is a marked increase towards the phantom edge for higher values of  $\sigma_t/L$, e.g. by a factor of 1.6 for  $\sigma_t/L$=50\,ps/m. This is because WEPL uncertainty depends on WEPL (see figure~\ref{fig:tof_physics_energy}) and in locations close to the edge, protons only traverse a thin tangential portion of the phantom under certain projection angles. The backprojection operation in the reconstruction therefore accumulates higher noise over all projections for points near the edge than for points close to the centre. A similar tendency is seen for the calorimeter-based system, where the 1\% case yields noise profiles comparable to the 30\,ps/m TOF system.  
At first sight, it seems surprising that the 50\,ps/m TOF case yields better precision than the 2\% calorimeter system at all distances from the centre although figure~\ref{fig:wepluncert_sigmatof} suggests similar performance. This is understandable because even at a location close to the edge within the cylinder, the proton WEPL is relatively large for most projection angles and the TOF system performs better than the calorimeter-based set-up for any WEPL larger than 8\,cm (figure~\ref{fig:tof_physics_energy}). 
 With the lower beam energy of 180\,MeV, which corresponds to 1.6\,cm residual range after 20\,cm of water, RSP precision reaches the intrinsic limit imposed by straggling and the beam's energy spread, at least for the inner half of the cylinder.

\section{Discussion}

The purpose of this work was to investigate the RSP precision in time-of-flight proton CT based on physics models and statistics. 
Our results show that an RSP precision of better than 1\% can be achieved in a water cylinder of 20\,cm and a beam energy of 200\,MeV, which is representative of a human head and therapeutic proton accelerator, and with a velocity error of a few tens of ps/m. The dose to the phantom centre  was 10\,mGy. 
Note that WEPL and thus RSP variance are inversely proportional to the square root of dose (equations~\ref{eq:sigmawepl} and~\ref{eqn:centraldose}) so that lowering the dose by a factor of 10 would increase (i.e. worsen) the RSP precision by a factor of  $\sqrt{10}\approx 3.14$. 
On the other hand, we point out that the RSP precision is inversely proportional to the fourth power of the pixel size, i.e. $\mathrm{Var}_\mathrm{RSP}(\vec{x}) \propto \Delta \xi^{-4}$ (equation~\ref{eq:Vp}), where a factor of $1/\Delta \xi^2$ results from the filtering operation and another factor of $1/\Delta \xi^2$ is linked to the number of protons per pixel (equation~\ref{eq:sigmawepl}) \parencite{Schulte2005,Quinones2016,Krah2020a}. 
Thus, increasing pixel size by a factor of two from $1\times1$\,mm$^2$ to $2\times2$\,mm$^2$ would reduce (i.e. improve) RSP precision by a factor of 4. 

WEPL uncertainty (and consequently RSP uncertainty) due to TOF measurement error decreases as the object thickness increases, while the opposite is true for energy straggling (figure~\ref{fig:wepluncert_sigmatof}). For a ratio of TOF error over distance between sensor planes of 10-20\,ps/m, both effects contribute to a similar extend. For velocity errors   $\sigma_t/L>50$\,ps/m, on the other hand, it is mainly the TOF measurement dominating the RSP precision. Our results show that a 30\,ps/m TOF system yields similar RSP precision with a cylindrical object of 20\,cm diameter as a 1\% precision calorimeter-based system (figure~\ref{fig:noiserecon}). 

The dependence of WEPL uncertainty on WEPL  leads to a spatially varying RSP precision in the reconstructed proton CT image. In particular, TOF proton CT images are expected to be noisier towards the object's edges than at its centre. This effect is not specific to TOF proton CT, however, but can also be seen in calorimeter-based proton CT \parencite{Radler2018,Dickmann2019}, where e.g. the energy measurement error in a scintillator varies as a function of deposited energy. \textcite{Bashkirov2016} studied this aspect and proposed a multilayer scintillator system in order to make WEPL uncertainty less dependent on WEPL. The design of a TOF proton CT system will likely require a similar kind of study to achieve homogeneous image noise. Possible ways to achieve this include energy degraders in front of the TOF system to reduce the residual energy, adapting the beam energy to the local phantom thickness \parencite{Dickmann2021}, or modulating the beam fluence to adjust the noise distribution \parencite{Dickmann2020}.

\textcite{Worstell2019} presented  a TOF proton CT prototype employing Large Area Picosecond Photon Detectors with a one sigma TOF error of 64\,ps. 
 In the setting used in this work, 
 i.e. a water cylinder of 20\,cm diameter, a dose to the centre of 10\,mGy, and $1\times1$\,mm$^2$ pixels,  these detectors are expected to yield similar RSP precision as a 2\% calorimeter-based proton CT system (figure~\ref{fig:noiserecon}) if the TOF sensors are placed 1\,m apart from each other. 
 With a spacing of $L=10$\,cm, as reported in \textcite{Worstell2019}, the uncertainty on the exit energy due to the TOF measurement error would increase by a factor of ten (equation~\ref{eq:sigmatof}) and the RSP precision would range from about 4\% at the centre to 13\% at the cylinder edge for a beam energy of 200\,MeV. 
Keeping the spacing at 10\,cm, but reducing the incident beam energy to 180\,MeV, which corresponds to a range in water of 21.6\,cm, the RSP at the cylinder centre would reduce to 0.9\% and to 9\% near the edge.  
 \textcite{Vignati2020} performed first tests in a therapeutic proton beam with a new ultra fast silicon detector based on low gain avalanche diode (LGAD) technology \parencite{Pellegrini2014}. They report a time resolution per sensor plane of 75\,ps to 115\,ps, i.e. TOF errors of 106\,ps to 162\,ps.  \textcite{Sadrozinski2018} had previously found TOF resolutions of an LGAD sensor of around 30\,ps.  
 \textcite{Curtoni2021} studied the performance of Chemical Vapor Deposition diamond detectors in the context of ion beam therapy monitoring and reported time resolutions on the order of 100-200\,ps for protons depending on the energy, and as low as 13\,ps for carbon ions. 
Based on our results, some of these new sensor types could be suitable for TOF proton CT in terms of time resolution. 
 In general, it would be crucial to devise an acquisition procedure which leads to an exit energy as low as possible (figure~\ref{fig:tof_physics_energy} left).  
Furthermore, given that the RSP resolution depends on the velocity error $\sigma_t/L$, designing a TOF proton CT will  also require engineering efforts to integrate into a treatment room  a system with a large distance between the sensor planes.

Our work makes a few simplifying assumptions. First, we only consider the energy measurement error, energy straggling, and the beam's energy spread as sources of RSP uncertainty, yet neglect noise induced by scattering in the object. \textcite{Dickmann2019} have studied the contribution of this source of noise in a calorimeter-based proton CT prototype \parencite{Johnson2016}. The prototype is designed in such a way that energy straggling and measurement error combined lead to a WEPL uncertainty roughly independent of WEPL \parencite{Bashkirov2016}.  
\textcite{Dickmann2019} found that noise due to scattering depends on the object shape and composition and the location of interest inside the object. The authors report that scattering noise in a homogeneous water cylinder is absent at the centre and most pronounced towards the cylinder edges where it contributes up to 20\% to the RSP precision. Based on this, our RSP precision profiles (figure~\ref{fig:noiserecon}, lower panels) are expected to remain unchanged at the centre and reach about 1\% for  $\sigma_t/L=$\,30\,ps/m if noise due to scattering is considered. 

Another simplification is that our reconstruction uses straight proton lines instead of curved most likely paths. This implies that our noise images (figure~\ref{fig:noiserecon}) are slightly blurrier than how they would be if we had implemented a reconstruction algorithm which considers the MLP, such as distance driven binning \parencite{Rit2013,Radler2018}. Given that RSP precision increases smoothly towards the edge, this simplification has no relevant impact on the results reported here. Note that using straight lines in the noise reconstruction and neglecting the contribution of MCS to image noise are two distinct aspects. The former causes the noise maps to be blurred while the latter may cause an underestimation of the noise level (see previous paragraph).

As the purpose of this work was to characterise a proton CT modality, we have only considered direct reconstruction algorithms, and in particular the distance driven method described in \textcite{Rit2013}. Other direct methods exist, but have been shown to yield very similar results in terms of image noise and spatial resolution \parencite{Khellaf2020}. We have not considered iterative reconstruction methods because image properties strongly depend on the parameters of the reconstruction, such as regularisation weights \parencite{Penfold2010,Penfold2015}.

\section{Conclusion}
The purpose of this work was to investigate the RSP precision in proton CT if TOF is used as energy measurement mechanism. 
The three noise contributions we considered were the TOF measurement error in relation to the spacing between the TOF sensor planes, the energy straggling experienced by the protons inside the scanned objects, as well as the proton beam's energy spread. For comparison, we also considered a calorimeter-based set-up which has been implemented in prototype scanners. 
Which of the three contributions dominates depends on the combination of beam energy and object thickness. For objects of up to 20\,cm and a beam energy of 200\,MeV, we find that beam spread and straggling are similarly relevant as the measurement error for velocity errors in the range of 10-50\,ps/m. In TOF proton CT systems with a distance between the sensor planes of one metre and a  time resolution of 100\,ps or larger, the measurement error tends to be the most dominant source of RSP uncertainty unless the beam energy is chosen such that the exit energy is very low (<80\,MeV). TOF proton CT images of a 20\,cm diameter water cylinder acquired at 10\,mGy total dose to the object centre and reconstructed with a pixel size of 1\,mm$^2$ have an RSP precision of 1\% or better when the velocity error is 50\,ps/m or lower. RSP precision varies spatially in a TOF proton CT image and generally increases towards the object's edges. This is similar to, but more pronounced than, previously reported results for calorimeter-based proton CT systems. A calorimeter-based system with 1\% energy error is comparable to a TOF-based system with a velocity error of 30\,ps/m, and a 2\% calorimeter system corresponds to a 64\,ps/m TOF system. 
Overall, we expect that better precision and more homogeneous image noise can be achieved by optimally adapting the beam energy to the phantom's water equivalent thickness.  

\section*{Acknowledgements}
The work of Nils Krah was supported by ITMO-Cancer (CLaRyS-UFT project). This work was performed within the framework of the LABEX PRIMES (ANR-11-LABX-0063) of Universit\'{e} de Lyon.

\section*{References}
\printbibliography[heading=none]

\end{document}